\documentclass[3p,times,procedia]{elsarticle}
\usepackage{nupha_ecrc}

\volume{00}

\firstpage{1}

\journalname{Nuclear Physics A}

\runauth{S. Gupta and R. Sharma}

\jid{nupha}

\jnltitlelogo{Nuclear Physics A}

\usepackage{amssymb}
\usepackage{slashed}

\newcommand\bef{\begin{figure}}
\newcommand\eef[1]{\label{fg:#1}\end{figure}}
\newcommand\beq{\begin{equation}}
\newcommand\eeq[1]{\label{#1}\end{equation}}
\newcommand\beqa{\begin{eqnarray}}
\newcommand\eeqa[1]{\label{#1}\end{eqnarray}}

\newcommand\fgn[1]{Figure \ref{fg:#1}}
\newcommand\eqn[1]{eq.\ (\ref{#1})}

\newcommand\tr{{\rm Tr\/}}
\newcommand\ppbar{\langle\overline\psi\psi\rangle}
\newcommand{\N}{{\cal N}}
\newcommand{\bilin}[1]{\overline\psi{#1}\psi}

\begin{document}

\begin{frontmatter}

\dochead{XXVIth International Conference on Ultrarelativistic Nucleus-Nucleus Collisions\\ (Quark Matter 2017)}

\title{Effective Field Theory Models for Thermal QCD}

\author{Sourendu Gupta and Rishi Sharma}
\address{Department of Theoretical Physics, Tata Institute of Fundamental Research, Homi Bhabha Road, Mumbai 400005, India.}
\address{}

\begin{abstract}
We present an effective field theory model for QCD at finite temperature
with quarks.  We discuss the mean field theory, the fixing
of parameters, and a prediction for
the curvature of the critical line. We proceed
to write down a pionic theory of fluctuations around the mean field,
and discuss how the parameters of this pionic effective theory descend
from the model with quarks.
\end{abstract}


\end{frontmatter}


\section{The effective action}\label{eft}

We build the model using the flavour symmetries SU$_V$($N_f$) $\times$
SU$_A$($N_f$), for $N_f$ flavours of quarks, acting on Fermion fields
which carry Dirac and flavour indices. We also carry along the SU($N_c$)
colour index, although these contribute only overall factors since there
are no colour interactions in the model: every fermion bilinear we use
is colour blind.  The dimension of the fermion field is $\N=4N_cN_f$
where $N_c=3$ and $N_f=2$.  The model is organized by the mass dimension
of operators, using an arbitrary temperature scale $T_0$ to adjust
dimensions if necessary.

We write an Euclidean finite temperature field theory using Hermitean
Euclidean Dirac matrices.  The Lorentz group becomes a rotation group
in Euclidean, with Hermitean generators $S_{\mu\nu} = -i[\gamma_\mu,
\gamma_\nu] /4$.  The theory breaks the full O(4) rotational symmetry down
to a cylidrical symmetry O(3)$\times$Z$_2$ of spatial rotational symmetry
and Euclidean time reversal symmetry T.  Every O(4) tensor also reduces,
time and space components of 4-vectors transform independently. Our
theory then has more couplings than a zero temperature theory.

There is always a possible dimension zero term in any EFT.  The
coefficient, $d^0 T_0^4$, called the vacuum energy, is set to zero.
There are no terms of dimension 1 or 2. The only allowed dimension
3 operator $\bilin{}$ has the quark pole mass as its coefficient,
which we write as $m_0=d^3T^0$. The allowed dimension 4 terms are
obtained by using derivative operators: $\bilin{\slashed\partial_4}$
and $\bilin{\slashed\partial}$.  Here $\slashed\partial_4 =
\gamma_4\partial_4$ and $\slashed\partial = \gamma_i\partial_i$.
The coefficient of the kinetic term, $\bilin{\slashed\partial_4}$ fixes
the normalization of the field operator, and hence is always set to
unity. The coefficient of the other term, $d^4$, is special to finite
temperature. It relates the pole mass to the quark screening mass.
All terms of dimension 5 can be eliminated by symmetry or on using the
equations of motion up to dimension 4. Similar arguments resctrict the
number of possible dimension 6 terms to ten.

The Euclidean EFT model we start with consists of all possible terms
up to mass dimension 6, invariant under the global and space-time symmetries
of a finite temperature Euclidean theory,
\beqa
  L &=& 
        d^3T_0\bilin{}
      + \bilin{\slashed\partial_4} 
      + d^4 \bilin{\slashed\partial_i} 
      + \frac{d^{61}}{T_0^2} \left[(\bilin{})^2 
            + (\bilin{i\gamma_5\tau^a})^2 \right]
      + \frac{d^{62}}{T_0^2} \left[(\bilin{\tau^a})^2 
            + (\bilin{i\gamma_5})^2 \right]\\
\nonumber &&
      + \frac{d^{63}}{T_0^2} (\bilin{\gamma_4})^2 
      + \frac{d^{64}}{T_0^2} (\bilin{i\gamma_i})^2
      + \frac{d^{65}}{T_0^2} (\bilin{\gamma_5\gamma_4})^2
      + \frac{d^{66}}{T_0^2} (\bilin{i\gamma_5\gamma_i})^2 
      + \frac{d^{67}}{T_0^2} \left[(\bilin{\gamma_4\tau^a})^2 
            + (\bilin{\gamma_5\gamma_4\tau^a})^2 \right]\\
\nonumber &&
      + \frac{d^{68}}{T_0^2} \left[(\bilin{i\gamma_i\tau^a})^2 
            + (\bilin{i\gamma_5\gamma_i\tau^a})^2\right]
      + \frac{d^{69}}{T_0^2} \left[(\bilin{iS_{i4}})^2 
            + (\bilin{S_{ij}\tau^a})^2 \right]
      + \frac{d^{60}}{T_0^2} \left[(\bilin{iS_{i4}\tau^a})^2 
            + (\bilin{S_{ij}})^2 \right]
\eeqa{eftaction}
This differs from the NJL model \cite{njl} in two important ways. First, Lorentz
invariance is given up since it is
built to model QCD at finite temperature, 
and a temperature scale $T_0$ is used to organize the expansion.
Second, it is an EFT, so all terms up to a certain order in mass dimension
are kept, provided they are invariant under the symmetries of the model.
The NJL model would have all four-fermi couplings set to zero except
$d^{61}$.

\section{The mean field approximation}\label{mft}

The mean-field approximation is the fermion operator identity
$\overline\psi_\alpha\psi_\beta = \delta_{\alpha\beta}\ppbar$,
where $\alpha$ and $\beta$ are composite Dirac-flavour-colour indices.
Performing the Wick-contractions in various ways in the generic 4-fermi term
then gives
\beq
 (\bilin{\Gamma})^2=2\ppbar\left[\tr\Gamma\bilin{\Gamma}
   -\bilin{\Gamma\Gamma}\right]
   -\ppbar^2\left[(\tr\Gamma)^2-\tr(\Gamma\Gamma)\right].
\eeq{fierz}
The product of Dirac-flavour matrices in the second term is the Fierz
transformation.  In this approximation, the EFT becomes
\beq
  L_{{\scriptscriptstyle{\rm MFT}}} = 
       -\N\left(\frac{T_0^2}{4\lambda}\right)\Sigma^2
      + \bilin{\slashed\partial_4} 
      + d^4 \bilin{\slashed\partial_i} + m\bilin{} 
\eeq{emft}
where $\lambda=(\N+2)d^{61}-2d^{62}-d^{63}+d^{64}+d^{65}-d^{66}
+d^{69}-d^{60}$, and we define $\Sigma=2\lambda\ppbar/T_0^2$,
and $m=d^3T_0+\Sigma$.  This is little more than the mean field
approximation to the NJL model, so we can utilize the body of work
in, for example, \cite{klevansky}. There are several interesting points
to note about the sum-integrals required to compute the free energy at
finite temperature. After performing the Matsubara sum, one can scale
the spatial momenta by $d^4$, as a result of which all thermodynamic
quantities contain $d^4$ as an overall factor. The integral for the free
energy density can be written in the form
\beq
 \Omega = \left(\frac{d^4}{2\pi}\right)^3 \int d^3p\;\left[
    TS_0(p) + E_0(p) + E(p,T)\right].
\eeq{formal}
The temperature independent pieces of the integrand have obvious
interpretations; $S_0(p)$ is a contribution to the entropy of the vacuum,
and $E_0(p)$ to its energy. These are formally divergent, and we choose
to use dimensional regularization to deal with them. Then the vacuum
entropy contribution vanishes and the vacuum energy is regulated. However,
this introduces a new scale, $M$, into the problem which can be called
the ``renormalization scale''. In the chiral limit, $d^3=0$, there
are two parameters to be determined ($d^4$ and $\lambda$) for a given
choice of $T_0$ and $M$. Due to the scaling of the free energy shown
in \eqn{formal}, one cannot use thermodynamic quantities to determine
the two parameters independently. Thermodynamics determines only the
combination $\lambda / (d^4)^3$.

\bef[htb]
\begin{center}
\includegraphics[scale=0.7]{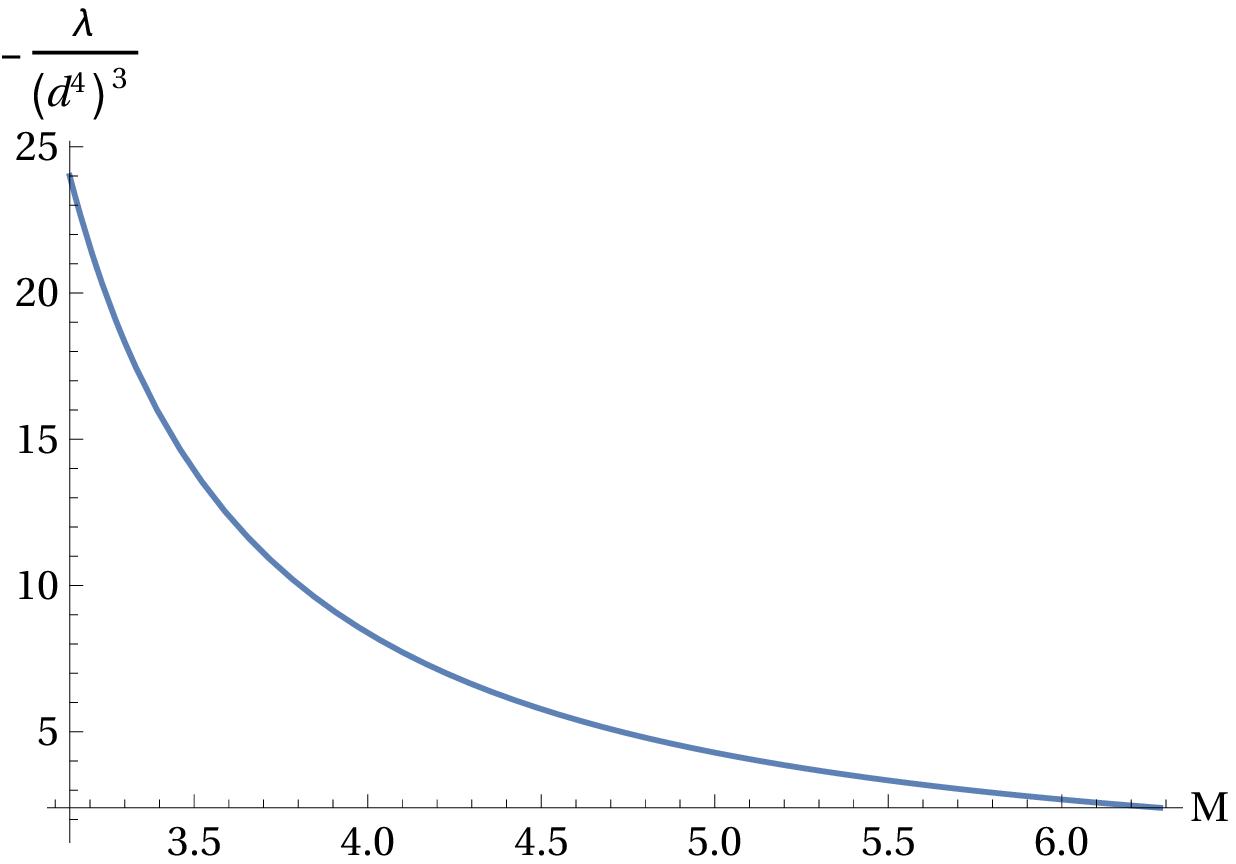}
\end{center}
\caption{The ``running'' of the quartic effective coupling as the
 ``renormalization scale'' $M$ (given in units of $T_c$) is changed
 in the chiral limit in order to keep $T_c$ fixed.}
\eef{lambdafit}

Since we are interested in the low-energy physics around the chiral
transition temperature, $T_c$, we choose $T_0$ to be $T_c$. This is
an arbitrary choice, and small changes in $T_0$ can be compensated by
changing the parameters of the Lagrangian, while keeping physical
quantities (such as the value of $T_c$) unchanged. Similarly,
changes in $M$ can be compensated by changes in the parameters. This
is the equivalent of renormalization-group flow for the effective
model. A typical running of the coupling with the scale is shown in
\fgn{lambdafit}. We determine the parameters of the model by taking
$T_c$ in the chiral limit to be a given value, and using the convention
$T_0=T_c$. It is natural to use a scale $M\simeq$500--700 MeV, between
the mass scales of the pseudoscalar and vector mesons. In this case
one can choose $\lambda/\N$ to be attractive and of order one for $T_c$
between about 140 Mev and 125 MeV. This is a technically natural value
of the coupling.

One interesting parameter-free observable emerges when we introduce a
chemical potential through a term $\tilde\mu T_0\bilin{\gamma_4}$. For
small $\mu$ the chiral critical point persists, albeit with a temperature,
$T_c(\tilde\mu)$ which shifts with the chemical potential.  The curvature
of the critical line in the chiral limit is usually given in terms of
the expansion $t_c(\tilde\mu) = 1 - T_c(0)\kappa \tilde \mu^2/2+{\cal
O}(\tilde\mu^3)$ where $t_c(\tilde\mu) = T_c(\tilde\mu)/T_c(0)$.
Estimates of this quantity have been made on the lattice with quarks
which are somewhat heavier than found in nature. The lattice results
correspond to the range $T_c(0)\kappa\simeq0.01$--0.05.

In the mean-field theory we find the completely parameter-free result
\beq
  t_c(\tilde\mu) =1-\frac3{2\pi^2}\tilde\mu^2, \qquad{\rm giving}\qquad
  T_c(0)\kappa=\frac3{\pi^2}.
\eeq{curvcrit}
Although the model prediction in the chiral limit is larger than the
lattice determinations, it is not too far away. Since the lattice results
are not in the chiral limit, and the model results are in the mean-field
approximation, it would be interesting to see how this prediction can
be refined.

\section{Fluctuations}

\bef[htb]
\begin{center}
\includegraphics[scale=1.0]{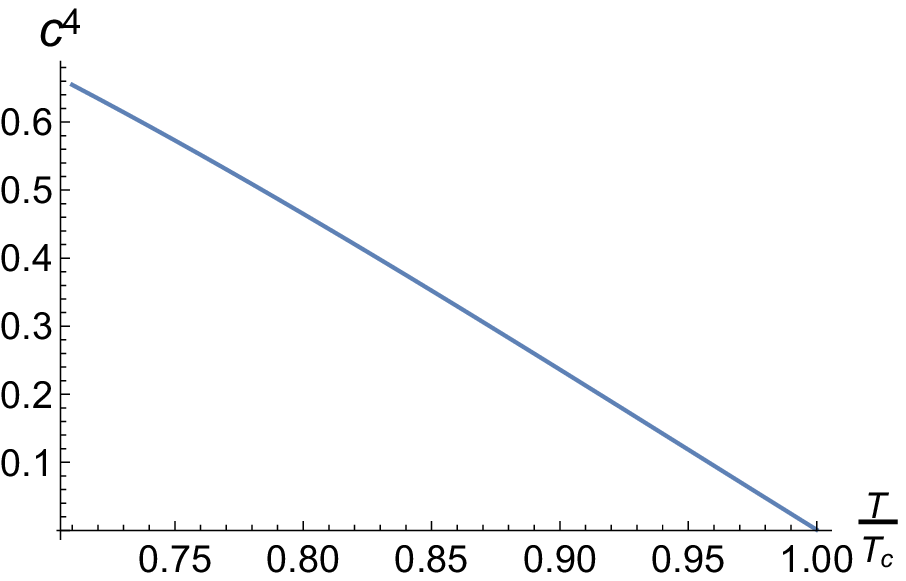}
\end{center}
\caption{The parameter $c^4$ computed by matching the quark effective
 theory to the pion effective theory in the chiral limit.}
\eef{c4chiral}

The mean-field effective action is invariant under the global vector SU(2)
transformations.  Local axial SU(2) transformations describe corrections
to the mean-field of the kind
\beq
  \overline\psi_\alpha\psi_\beta = \ppbar U_{\alpha\beta}
     \qquad{\rm where}\qquad
   U=\exp\left(\frac{i\tilde\pi\gamma_5}f\right)\times1,
\eeq{fluct}
where $U$ acts in flavour space, and is the identity in Dirac-colour space.
The Lagrangian for quadratic fluctuations of coupled to quarks in the mean
field approximation is then
\beq
  L = -\frac{\N}{4\lambda}\Sigma^2
      +\bilin{\left\{m+\frac{im_0\tilde\pi\gamma_5}f-\frac{m_0\tilde\pi^2}{f^2}\right\}}
      +\bilin{\left\{\slashed\partial-\frac i{2f}\gamma_5\slashed\partial\tilde\pi
         -\frac1{8f^2}[\slashed\partial\tilde\pi,\tilde\pi]\right\}}.
\eeq{pionlag}
The quarks have to integrated out in order to get the effects of the
fluctuations. The terms linear in pion fields give tadpole contributions.
However, these vanish due to the trace over isospin, $\tau^a$. The 
triangle diagram giving the dimension 3 pion self coupling also vanishes.
As a result, the effective action for fluctuations is quadratic
\beq
  L_f = \frac{c^2 T_0^2}2\pi^2 + \frac12(\partial_0\pi)^2 
    + \frac{c^4}2(\nabla\pi)^2 
\eeq{quadpi}
where the field $\pi=\sqrt{Z_\pi}\tilde\pi$. $Z_\pi$, $c^4$ and $c^2$
can be computed from \eqn{pionlag} by examining two-point function of
the pion and then integrating over the quark fields. Such a quadratic
theory has been examined earlier \cite{sonsteph}. In our approach, this
requires constructing an IR effective field theory of pions by matching
its parameters to that of the UV effective theory of quarks.

This matching gives us a relation for the effective pion pole mass which
is similar to the Gell-Mann-Oakes-Renner relation,
\beq
 c^2T_0^2 = -\,\frac{m_0\ppbar}{f^2Z_\pi}
\eeq{gor}
However, unlike the relation at $T=0$, we have not yet made an
identification of $f$ with a measurable scale.  So the content of the
equation above, as yet, is to define $f$ in terms of other computable
quantities.  The remaining parameters of the model, namely $c^4$ and
$Z_\pi/f^2$ can also be determined. 

In the chiral limit we find that $c^4$ vanishes linearly as $T\to T_c$
(see, for example, \fgn{c4chiral}).
As a result, the quadratic theory fails at the critical point, and it
is necessary to compute higher order terms in the effective theory in
order to obtain critical properties. 

Further investigations, including
that of the parameters of the effective pion theory away from the
chiral limit, are on.

SG would like to acknowledge his J.~C.~Bose grant number SR/S2/JCB-100/2011.






\end{document}